\documentclass[fleqn,12pt,twoside]{article}
\usepackage{espcrc1}

\usepackage{graphicx}
\usepackage[figuresright]{rotating}

\newcommand{\AmS}{{\protect\the\textfont2
  A\kern-.1667em\lower.5ex\hbox{M}\kern-.125emS}}

\newcommand {\beq}{\begin{eqnarray}}
\newcommand {\eeq}{\end{eqnarray}}
\newcommand {\be}{\begin{equation}}
\newcommand {\ee}{\end{equation}}

\def\bfm#1{\mbox{\boldmath $#1$}}
\newcommand{\BQ}{\begin{equation}}
\newcommand{\EQ}{\end{equation}}
\newcommand{\BQA}{\begin{eqnarray}}
\newcommand{\EQA}{\end{eqnarray}}

\newcommand{\kv}{\mbox{\boldmath $k$}}

\newcommand{\q}{\mbox{\boldmath $q$}}
\newcommand{\p}{\mbox{\boldmath $p$}}

\title{
Structure change of Cooper pairs in color superconductivity \footnote{
Based on the work with H.~Abuki and T.~Hatsuda \cite{AHI} and the 
on-going collaboration with G.~Baym and T.~Hatsuda \cite{BHI}.}\\
       --- Crossover from BCS to BEC ? ---}

\author{Kazunori Itakura \address{RIKEN BNL Research Center, 
BNL, 
              Upton, NY 11973, USA}}
       
\begin{document}

\maketitle

\begin{abstract}

We discuss a possibility of transition from color superconductivity 
of the standard BCS type at high density, to Bose-Einstein Condensation
(BEC) of 
Cooper pairs at lower density. Examining two-flavor QCD
over a wide range of baryon density, we found the size of a Cooper pair  
becomes small enough to be comparable to the averaged quark-quark 
distance at lower density. We also consider the same problem in two-color QCD.

\end{abstract}

\section{Introduction}
There are two reasons for expecting that color superconductivity at 
moderate density could be qualitatively different from the usual 
weak-coupling superconductivity in metal.  
They are both related to distinct properties of QCD. 
As usual, Cooper instability is induced by quark-quark attractive 
interaction, but what is different in quark matter is that it is 
effective in principle for {\it all of the quarks} inside the Fermi sea.
This is because the interaction itself has an attractive channel due to 
color factor. This property is in clear contrast to the electron 
superconductivity where the Coulomb interaction is repulsive, and 
the attractive force by the phonon exchange exists only in a small 
region $|\epsilon_k-\epsilon_{\rm F}|<\omega_{\rm D}$
restricted by the Debye cutoff $\omega_{\rm D}$. 
At high baryon density, the attractive force is given by 
one gluon exchange and becomes weak because the coupling constant 
at typical momentum scale (the chemical potential $\mu$) 
becomes small due to the asymptotic freedom. This ensures weak-coupling 
treatment of the color superconductivity, and much of efforts has been 
done in this direction \cite{Review}. 
However, this implies, at lower densities, the 
effective attractive interaction becomes large, which will invalidate  
the naive weak-coupling treatment. 
These two properties (absence of analogue of $\omega_{\rm D}$ and the 
infrared enhancement of the QCD coupling) suggest that as the 
density becomes low, Cooper pairing will not be restricted only 
to a vicinity of the Fermi surface and become much more drastic phenomena. 
Besides, the Cooper pair itself 
will change into a tightly bound state with small size. Therefore, 
it is quite natural to expect that at low density (but still above the 
critical density of the deconfinement transition) the color superconductivity 
will turn into ``strong-coupling superconductivity'' or even 
``Bose-Einstein Condensation'' of tightly bound Cooper pairs. 
To investigate this possibility is our main purpose of this talk. 
In the following, we will first discuss two-flavor case 
over a wide range of baryon density with a single model \cite{AHI}.
Structural change of a Cooper pair is best studied by computing its  
wavefunction (quark correlation in the color superconductor) which is 
easily obtained once we know the momentum dependence of a superconducting gap. 
We will also consider the same problem in the two-color case \cite{BHI}.



\section{Gap equation}
A common field-theoretic strategy of superconductivity is 
the Nambu-Gor'kov formalism which uses a two component Dirac spinor
$
\Psi = \left(
\matrix{\psi \cr \psi^c\cr}\right),\  \psi^c=C\bar\psi^T.  
$
The extended Fermion propagator $i S(x-y)
\equiv\langle T \Psi(x)\bar\Psi(y)\rangle $ is now a $2\times 2$ matrix. 
The Schwinger-Dyson equation for self energy $\Sigma = S_0^{-1}-S^{-1}$
is written as
\BQ
\Sigma(k) =  \left(\matrix{
M(k)& {\Delta(k)}\cr
\gamma^0\Delta(k)^\dag \gamma^0& M(-k)
}\right) =  \int  \frac{d^4q}{(2\pi)^4}\ g^2\ {\Gamma_\mu^a} 
S(q)\ \Gamma^a_\nu \ 
D^{\mu\nu} (k-q), \label{SD}
\EQ
where we ignore quark mass, $D^{ab}_{\mu\nu}=\delta^{ab}D_{\mu\nu}$ 
is the gluon propagator in medium (which includes Debye screening for 
electric gluons and 
Landau damping for magnetic), $S(q)$ is the full quark propagator, and
$\Gamma^a_\mu$ is the quark-gluon vertex, which is taken to be a bare one 
$\Gamma_\mu^a={\rm diag}(\gamma_\mu T^a,-\gamma_\mu (T^a)^T )$.
For $g^2$ in eq.~(\ref{SD}), we use a momentum dependent coupling 
$g^2(q,k)$ in the  ``improved ladder approximation'' \cite{Higashijima} 
{\bf (}$\beta_0=(11N_c-2N_f)/3${\bf )}:
\beq
  g^2(q,k)=\frac{16\pi^2}{\beta_0}\, \frac{1}{\ln \left(%
  (p^2_{\rm max} +p_c^2)/{\Lambda^2} \right)},
  \quad p_{\rm max}={\rm max}(q,k),
  \label{HM}
\eeq
where $p_c^2$ plays a role of a phenomenological 
infrared regulator. At high momentum, 
  $g^2$ shows the same logarithmic behavior as the usual running coupling 
  with $\Lambda$ identified with $\Lambda_{\rm QCD}$.
We adopt $\Lambda$=400 MeV and $p_c^2$=1.5 $\Lambda^2$ which 
 are determined to reproduce the low energy meson properties for 
  $N_f=2, N_c=3$ vacuum.

Performing the angular and frequency integrals leads 
to a gap equation with momentum dependence only. 
Once we obtain the momentum dependent gap $\Delta(q)$, we can compute the 
wavefunction of a Cooper pair (or, $q$-$q$ correlation function) in 
momentum space:
\BQ
{\varphi}(q)=  
\frac{\Delta(q)}{2\sqrt{(q-\mu)^2 +|\Delta (q)|^2}}. 
\EQ
The size of Cooper pairs (the coherence length) is defined as the root 
mean square radius of coordinate space wavefunction $\tilde\varphi(r)$. 
Recall that in a typical type-I superconductor in metals, 
the size of a Cooper pair $\sim \Delta^{-1}$ is 
much larger than the typical scale 
$\sim k_{\rm F}^{-1}$ (the ratio is $k_{\rm F}/\Delta \sim 10^{4}$), because 
there is a clear scale  hierarchy, $\Delta \ll \omega_{\rm D} \ll k_{\rm F}$.
On the other hand, since there is no intrinsic scale $\omega_{\rm D}$ 
in QCD, scale hierarchy at extremely high density simply reads 
$\Delta\sim \mu e^{-c/g} \ll k_{\rm F}\sim \mu$.  
 At lower densities, however,  such scale separation becomes
    questionable for $g$ is not small.

\section{Momentum dependent gap and size of a Cooper pair in $N_f=2, N_c=3$ 
\cite{AHI}}

When $N_f=2$ and $N_c=3$, 
the most attractive channel is the 
flavor anti-symmetric, color anti-symmetric and $J=0^+$ channel
$
\Delta(k)=(\tau_2^{\rm (flavor)}\lambda_2^{\rm (color)}  C\gamma_5)
 \{ \Delta_+(k)\Lambda_+(\hat{\kv} ) +
    \Delta_-(k)\Lambda_-(\hat{\kv} ) \} ,
$
where $\tau_2$ is the Pauli matrix acting on the flavor space, 
$\lambda_2$ is a Gell-Mann matrix, and  $C$ is the charge conjugation. 
$\Lambda_\pm(\hat{\kv})\equiv%
(1\pm\hat{\kv}\cdot\bfm{\alpha})/2$ is the projector on 
 positive ($+$) and negative $(-)$ energy quarks.
We ignore the effects of chiral condensate.   

In Fig.~1(a), we show the gap $\Delta_+(k=|\kv|)$ as a solution of
   the gap equation  for a wide range of densities.
This result tells us the following things. 
At {\bf high density}, 
  (i) {\it there is a  sharp peak at the Fermi surface},
      and 
  (ii) {\it the gap decays rapidly but is nonzero for momentum far away 
   from the Fermi surface}.
 The property (i) is similar to the standard BCS
 but (ii) is not, which is due to the absence of the 
 Debye cutoff in the gluonic interaction.
On the other hand, at {\bf low density},  
 (iii) {\it the sharp peak at the Fermi surface disappears}, and
 (iv) 
{\it color superconductivity
 at low density is not a phenomenon just around the Fermi surface}. 
 This last point can be said after a close look at the effects of each 
 contribution to the gap and computing the occupation number. 
 The result shows that the Fermi surface is diffuse 
 substantially at low density \cite{AHI}. 

Computing the size of a Cooper pair, 
   one finds that it is less than 4 fm at the lowest density 
   shown in Fig.~1 ($\mu=800$MeV) \cite{AHI}.
This is small enough from the usual sense, because the  
ratio $\xi_{\rm c}/d_q$ of the Cooper pair size $\xi_c$ to 
typical length of the system $d_q$, i.e., 
averaged inter-quark distance  for free quarks 
$d_q=({\pi^2}/{2})^{1/3}/\mu$, is 
less than 10, in contrast to $10^5$ at the highest density (Fig.~1(b)). 
If we further extrapolate the curve in Fig.~1(b) to lower chemical potential
  and are still in the deconfined phase,
  tightly bound Cooper pairs ($\xi_{\rm c}/d_q\sim 1$) may seem to appear. 
Therefore, loosely bound large Cooper pairs 
   similar to the BCS superconductivity in metals are formed
   at extremely high density, while at lower density, the size of a 
   Cooper pair is small enough.
 This smooth transition from  $\xi_{\rm c}/d_q \gg 1 $ to 
  $\xi_{\rm c}/d_q \sim 1$ as $\mu$ decreases is analogous to the 
  crossover from the BCS-type
  superconductor to the BEC of tightly bound Cooper pairs 
    \cite{BEC}.

\begin{figure}[htbp]
 \vspace{-2mm}
 \begin{minipage}{0.49\textwidth}
 \includegraphics[scale=0.35]{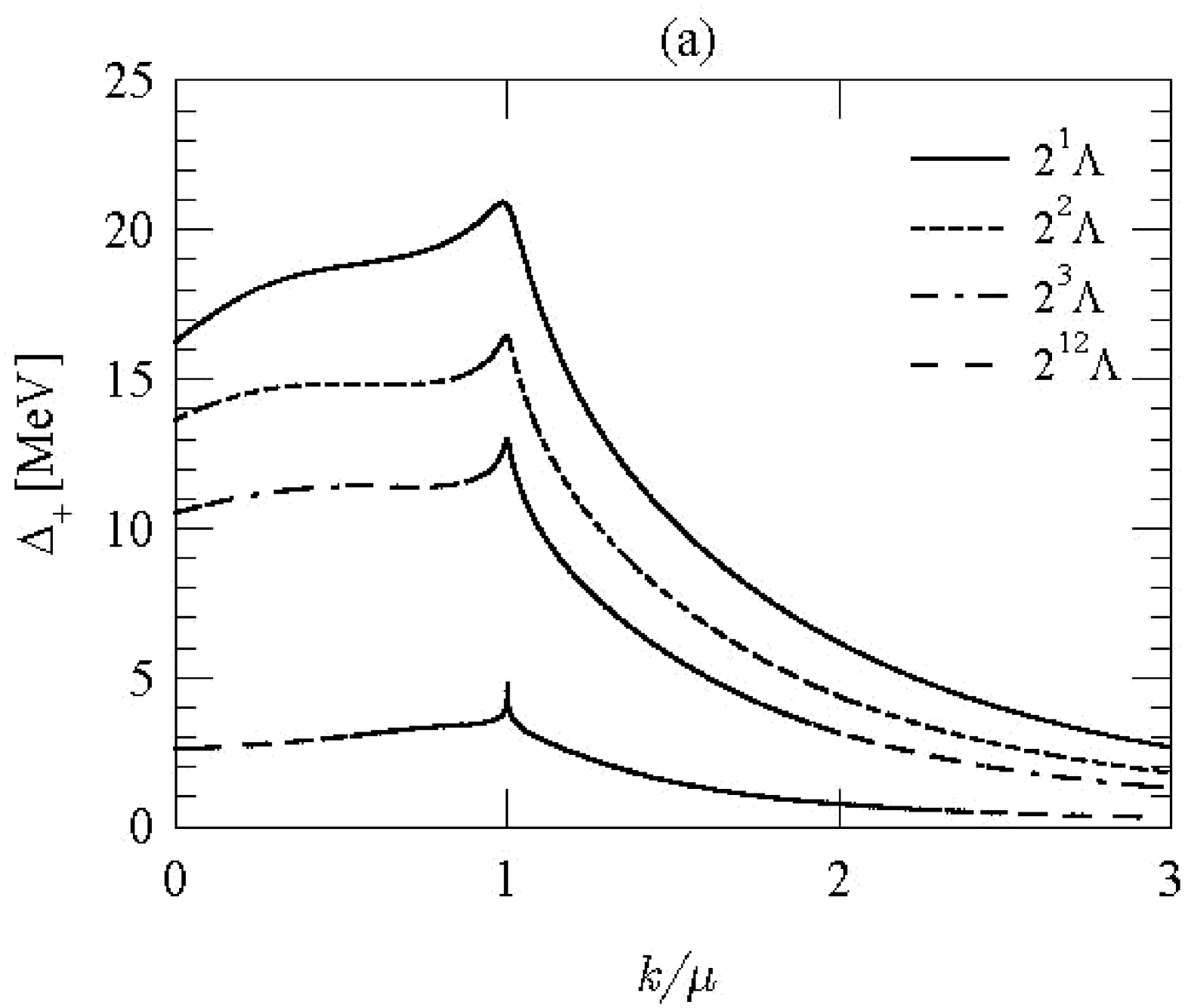} 
 \end{minipage}%
 \hfill%
 \begin{minipage}{0.49\textwidth}
  \includegraphics[scale=0.49]{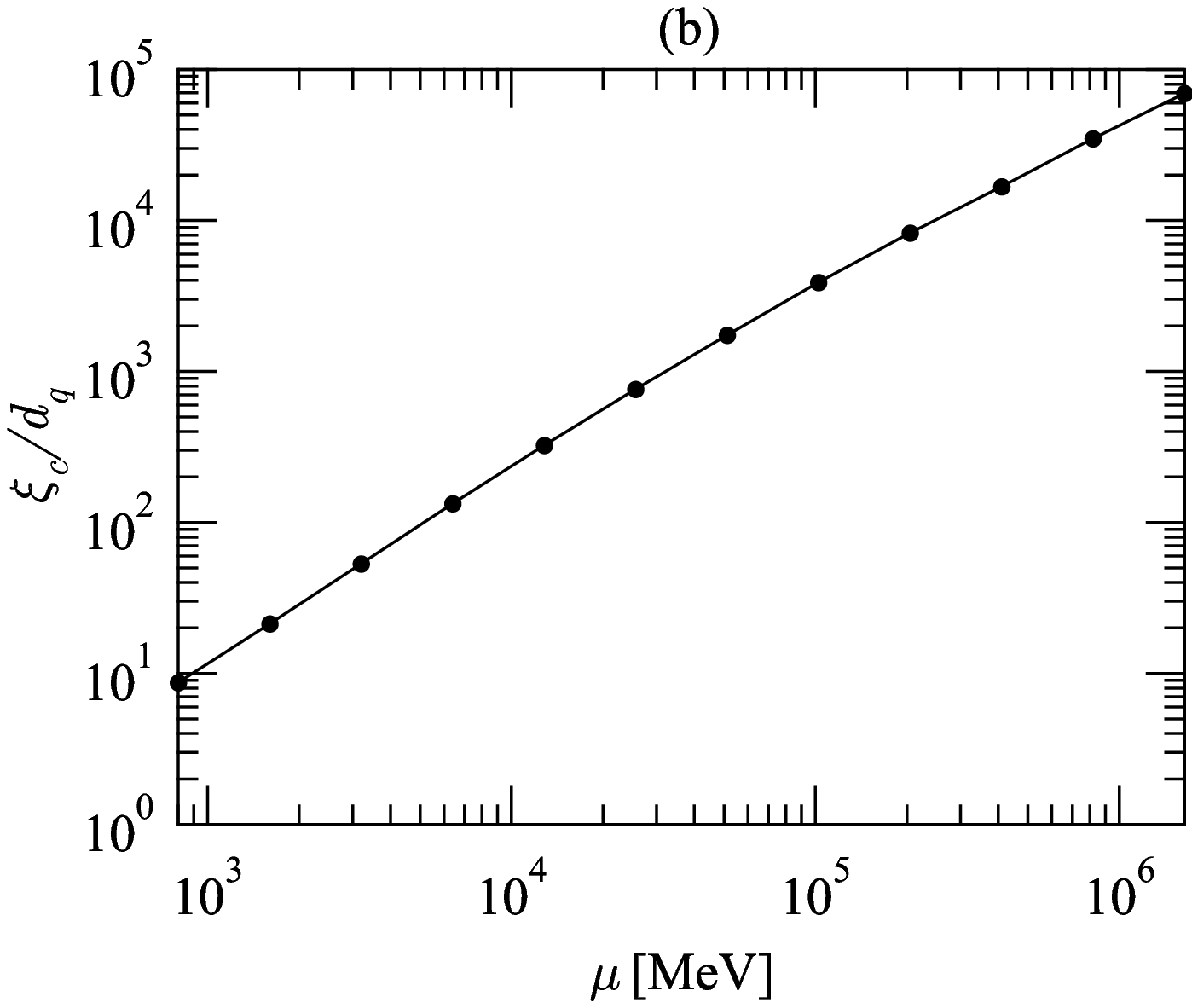} 
 \end{minipage}%
 \vspace{-8mm}
  \caption{(a): $\Delta_+(k)$ as a function of $k/\mu$ for various densities
        $\mu =2^n\Lambda $ with $n=1,2,3,12$.
    (b): Ratio of the coherence length to the
          average inter-quark distance as a function of $\mu$. 
          }
\vspace{-8mm}
\end{figure}

\section{Two color QCD with two flavors \cite{BHI}}

It is interesting to study the same problem the same way in 2-color QCD 
\cite{BHI}, 
where the diquark $\Delta(p) = (\tau_2^{\rm (color)}\tau_2^{\rm (flavor)} 
\gamma_5) \Delta_s(p)$ is a gauge invariant ``baryon'' and  
at low density, it corresponds to the Nambu-Goldstone (NG) boson of a 
spontaneous broken enlarged chiral symmetry group \cite{2color}. 
Therefore, bosonic-like description of diquark is 
expected to be appropriate at least at low density. 
The formalism developed in Sect.~2 is directly applicable to this problem, 
but we here retain the chiral condensate $M\neq 0$. 
Then the Schwinger-Dyson equation (\ref{SD}) forms coupled equations for 
$M(p)$ and $\Delta(p)$: 
\BQA
M(p)\!\!\!&=&\!\!\!\frac{3}{8}\, g^2 \int \frac{d^3\q}{(2\pi)^3}\  D_{\nu\nu}(\p-\q) \ 
\frac{M(q)}{\sqrt{q^2+M(q)^2}}\ \Big(1-n_-(q)-n_+(q)\Big),\\
\Delta(p)\!\!\!&=&\!\!\!\frac{3}{8}\, g^2 \int \frac{d^3\q}{(2\pi)^3}\  D_{\nu\nu}(\p-\q) \ 
\frac12 \left[\frac{1}{\epsilon_-(q)}+\frac{1}{\epsilon_+(q)}\right]\Delta(q)
\EQA
where $\epsilon_\pm(q) = \sqrt{(E\mp \mu)^2+\Delta^2}$ 
is the quasi-particle energy (with $E(q)=\sqrt{q^2+M^2(q)}$) and 
$n_\pm = \{1-(E\pm \mu)/\epsilon_\pm\}/2$ is the occupation number.
It is important to notice that,  in the limit $\mu\to 0$,  these gap 
equations have the same form
which is invariant under the mixture of vector $(M(p),\Delta(p))$: 
\BQ
\pmatrix{M(p)\cr\Delta(p)}
\approx \frac{3}{8}\, g^2 \int \frac{d^3\q}{(2\pi)^3} D_{\nu\nu}(\p-\q) 
\frac{1}{\sqrt{q^2+M^2+\Delta^2}} \pmatrix{M(q)\cr\Delta(q)} \qquad (\mu\to 0).
\EQ
This observation is consistent with the Pauli-G\"ursey symmetry which 
is present at $\mu=0$ and leads to a strong consequence. 
For simplicity, we consider the chiral limit.  
We know that, at $\mu=0$, chiral symmetry is broken $M\equiv 
M_0\neq 0$ 
and the diquark condensate is zero $\Delta=0$ (there is no Fermi sphere). 
Consequently,  
a tightly $q\bar q$ bound state ``pion'' is generated as a NG boson, i.e.,  
``the NG pion''. On the other hand, if the chemical potential is slightly 
nonzero $\mu\neq 0$, the symmetry in eq.~(6) is weakly broken so that 
the nonzero diquark condensate $\Delta\simeq M_0\neq 0$ is favored 
and the chiral condensate is zero $M=0$. The associated 
NG boson corresponds to a tightly-bound diquark state ``the NG baryon''.
Therefore, low density corresponds to the BEC-like region. 
If the quark mass is nonzero, the exchange between $(M\neq 0,\, \Delta=0, 
{\rm \, NG\ pion})$ at $\mu=0$ and $(M= 0,\, \Delta\neq 0, 
{\rm \, NG\ baryon})$ at $\mu\neq 0$ occurs smoothly. 
On the other hand, at very high density, we can ignore the chiral 
condensate, and the gap equation for $\Delta(p)$
becomes the same as the previous 3 color case up to the prefactor. 
Thus, as the density is increased, the bosonic-like description should be 
replaced by the usual BCS-type description. 
There must be a transition from BEC to BCS as the density is increased. 
More evidences of this transition are now under investigation \cite{BHI}.


\begin{thebibliography}{0}
\bibitem{AHI}
        H.Abuki, T.Hatsuda and K.Itakura, 
        Phys.Rev.{\bf D65},\, 074014 (2002); hep-ph/0206043.   
\bibitem{BHI}
        G. Baym, T. Hatsuda and K. Itakura, work in progress.
\bibitem{Review}
        For a recent review, see
        K. Rajagopal and F. Wilczek, ``{\it The Condensed Matter Physics
        of QCD}'', hep-ph/0011333.
\bibitem{Higashijima}
        K. Higashijima, Phys. Rev. {\bf D29}, 1228 (1984);
        Prog. Theor. Phys. Suppl. {\bf 104}, 1 (1991).  
         V. A. Miransky, Sov. J. Nucl. Phys. {\bf 38}, 280 (1983).    
\bibitem{BEC}
        P. Nozi\'eres and S. Schmitt-Rink, J. Low Temp. Phys. {\bf 59},
        195 (1985). \\
        See also, 
        E. Babaev, Int. J. Mod. Phys. {\bf A16}, 1175 (2001)
        and references therein. 
\bibitem{2color}
        T. Sch\"afer, E.V.Shuryak, M.Velkovsky, Phys. Rev.Lett. 81, 53 (1998);
        J.B.Kogut, M.A.Stephanov, and D. Toublan, Phys. Lett. {\bf B464},
        183 (1999); J.B.Kogut, et al., Nucl. Phys. {\bf B582}, 477 (2000).  
 \end{thebibliography}
\end{document}